\documentclass[12pt]{iopart}
\usepackage[dvipdfmx]{graphicx}
\usepackage{amsfonts}
\usepackage[subrefformat=parens]{subcaption}
\usepackage{color}
\usepackage{braket}
\usepackage{hyperref}
\usepackage{times}
\usepackage{braket}
\usepackage{ulem}
\usepackage{bm}
\usepackage{enumerate}
\usepackage{color}

\begin{document}

\title[]{Preparing ground states of the XXZ model using the quantum annealing with inductively coupled superconducting flux qubits}

\author{Takashi Imoto}
\address{Research Center for Emerging Computing Technologies, National Institute of Advanced Industrial Science and Technology (AIST),
1-1-1 Umezono, Tsukuba, Ibaraki 305-8568, Japan.}

\author{Yuya Seki}%
\address{Graduate School of Science and Technology, Keio University, Hiyoshi 3-14-1, Kohoku-ku, Yokohama 223-8522, Japan.}

\author{Yuichiro Matsuzaki}
\address{Research Center for Emerging Computing Technologies, National Institute of Advanced Industrial Science and Technology (AIST),
1-1-1 Umezono, Tsukuba, Ibaraki 305-8568, Japan.
}

\ead{matsuzaki.yuichiro@aist.go.jp}

\vspace{10pt}

\begin{abstract}
Preparing ground states of Hamiltonians is important in the condensed matter physics and the quantum chemistry.
The interaction Hamiltonians typically contain not only diagonal but also off-diagonal elements.
Although quantum annealing provides a way to prepare a ground state of a Hamiltonian,
we can only use the Hamiltonian with Ising interaction by using currently available commercial quantum annealing devices.
In this work, we propose a quantum annealing for the XXZ model, which contains both Ising interaction and energy-exchange interaction, by using inductively coupled superconducting flux qubits.
The key idea is to use a recently proposed spin-lock quantum annealing where the qubits are driven by microwave fields. As long as the rotating wave approximation is valid, the inductive coupling between the superconducting flux qubits produces the desired Hamiltonian in the rotating frame, and we can use such an interaction for the quantum annealing while the microwave fields driving play a role of the transverse fields.
To quantify the performance of our scheme, we implement numerical simulations, and show that we can prepare ground states of the two-dimensional Heisenberg model with a high fidelity.
\end{abstract}

%
%
%
%
%

\section{Introduction}

Quantum annealing (QA) is a method to solve the combinatorial optimization problem~\cite{kadowaki1998quantum,farhi2000quantum,farhi2001quantum}.
Combinatorial optimization problems can be mapped into problems of finding the ground state of an Ising model.
In this case, the Ising model is the problem Hamiltonian.
On the other hand, we use the Hamiltonian of transverse fields to prepare an initial state, and this is called a drive Hamiltonian.
In the QA, we prepare a ground state of the drive Hamiltonian, and then let the system evolve by a time dependent Hamiltonian to change from the drive Hamiltonian to the problem Hamiltonian in an adiabatic way.
This provides a way to prepare a ground-state of the problem Hamiltonian as long as an adiabatic condition is satisfied.

There are many other applications of the quantum annealing in addition to combinatorial optimization problems.
Firstly, quantum annealing can be used for machine learning.
A way to learn binary machine learning models using the quantum annealing is reported~\cite{date2020adiabatic, sasdelli2021quantum}. 
There are applications of the quantum annealing for clustering~\cite{kurihara2009quantum, kumar2018quantum}.
A method of using the quantum annealing to perform calculations for the topological data analysis(TDA) was reported~\cite{berwald2018computing}.
Quantum annealing has been applied as a sampling machine for restricted Boltzmann machines with deep layers~\cite{adachi2015application, wilson2021quantum, li2020limitations, neven2008training, winci2020path}.
Secondly,
quantum annealing is known to be useful for quantum chemistry calculation.
The Hamiltonian of molecules are written in the second quantized form.
There are known methods to convert 
the second quantized form of the Hamiltonian into the spin Hamiltonian~\cite{bravyi2002fermionic, verstraete2005mapping, seeley2012bravyi, tranter2015b, xia2017electronic}.
In quantum chemical calculations, it is important to obtain the energy of 
the molecules with high accuracy.
When the energy is known to the so called chemical accuracy, a chemical reaction can be predicted~\cite{eyring1935activated}.
In fact, calculations to
determine the energy of molecules
using quantum annealing have been reported~\cite{copenhaver2021using,  mazzola2017quantum, genin2019quantum, streif2019solving}.
In addition, the method to measure similarity of the structures of molecules
using the quantum annealing is proposed~\cite{hernandez2017enhancing}.
Finally, the quantum annealing can be applied to
simulate condensed matter phenomena
The $\mathbb{Z}_{2}$ spin liquid is emulated with the quantum annealing~\cite{zhou2021experimental}.
It has been reported that the KT(Kosterlitz–Thouless) phase transition was observed using a quantum annealing on the d-wave system~\cite{king2018observation}.
Simulations of the Shastry-Sutherland Ising, a frustrated system, were performed using quantum annealing~\cite{kairys2020simulating}.
A phase transition of three dimensional transverse Ising model is detected using the quantum annealing machine~\cite{harris2018phase}.
It is worth mentioning that, in the previous demonstrations, the problem Hamiltonian was mapped into an Ising model, and this typically requires many ancillary qubits.

Superconducting flux qubits (FQs)
are considered as promising systems for quantum information processing. The FQ is composed of a superconducting loop with Josephson junctions~\cite{mooij1999josephson, orlando1999superconducting, clarke2008superconducting, makhlin2001quantum}.
It is possible to tune the energy gap of the FQ and the interaction between FQs~\cite{paauw2009tuning, niskanen2007quantum}. There are many attempts to improve the coherence time of the FQs~\cite{burkard2005asymmetry, yoshihara2006decoherence, bylander2011noise}.
Moreover, recently, a new design of the FQs called a capacitivly shunted flux qubit with a long coherence time was proposed and demonstrated~\cite{you2007low, yan2016flux, corcoles2011protecting, steffen2010high}.
The FQ can be used as a sensor to detect magnetic fields, electron spins, or defects \cite{bal2012ultrasensitive,toida2019electron,budoyo2020electron,abdurakhimov2020driven}.
The FQs have advantages in a scalability, and actually
it is possible to fabricate thousands of the FQs even in the current technology
\cite{kakuyanagi2016observation}.

The company D-Wave Systems, Inc. realizes 
a quantum annealer with thousands of the FQs~\cite{johnson2011quantum, boixo2014evidence}.
The energy bias, energy gap, and
the coupling between the FQs can be tuned by changing the applied magnetic flux. 
With currently available commercial
quantum annealing devices,  we can use only transverse field Ising model as the Hamiltonian where the flux qubits are inductively coupled.
Recently, two flux qubits are coupled by using two degree of freedoms, charge and
flux~\cite{ozfidan2020demonstration}. In this case, not only Ising coupling but also energy-exchange interaction such as $\hat{\sigma}_x\hat{\sigma}_x$ (or $\hat{\sigma}_y\hat{\sigma}_y$) can be realized.
However, this approach requires a complicated setup. It is preferable if we could construct the energy-exchange terms with less degree of freedom of the FQs.

In this paper, we propose a method to prepare a ground-state of the Hamiltonian that contains both Ising interaction and energy exchange
interaction, which is the XXZ model, by using QA with the inductively coupled FQs.
The key idea is to adopt a spin lock quantum annealing~\cite{chen2011experimental,nakahara2013lectures,matsuzaki2020quantum}.
When we drive the FQs with the microwave fields, we can rewrite the Hamiltonian in a rotating frame with a frequency of the microwave fields. By using a rotating wave approximation, the inductive coupling between the FQs provides not only Ising interaction but also the energy exchange interaction. The driving with the microwave fields play a role of the driver Hamiltonian of the QA, and we can obtain a ground state of a Hamiltonian that contains both Ising and energy-exchange interaction by adiabatically turning off the driving fields.
Such a preparation of the ground state allows us to calculate an arbitrary correlation function for the ground state of the XXZ model, which is useful in the condensed matter physics~\cite{sachdev2011quantum}.
To  quantify  the  performance  of our  scheme,  we  implement  numerical  simulations to prepare  ground  states of the  two-dimensional  Heisenberg  model, and shows that we can achieve a high fidelity.

The paper is structured as follow.
In Section \ref{sec:prev_work}, we review the conventional quantum annealing and spin lock quantum annealing.
In Section \ref{sec:method}, we show the method to 
create the energy exchange interaction by using the inductivly coupled FQs for the QA.
In Section \ref{sec:numerical_simulation}, we perform numerical simulations to evaluate the performance of our method.
In section \ref{sec:conclusion}, we summarize and discuss our results.


\section{The previous work}\label{sec:prev_work}

In this section, let us
review the conventional quantum annealing with DC fields and recently proposed quantum annealing with spin-lock technique.

\subsection{Quantum annealing}

In this subsection, we review the conventional quantum annealing.
The annealing Hamiltonian is written as
\begin{equation}
    H(t)=\frac{t}{T}H_{P}+\biggl(1-\frac{t}{T}\biggr)H_{D}
\end{equation}
where $T$ is annealing time, $H_{P}$ is the problem Hamiltonian, and $H_{D}$ is the drive Hamiltonian.
We define the drive Hamiltonian as follows
\begin{equation}
    H_{D}=-B\sum_{j=1}^{L}\hat{\sigma}_{j}^{(x)}
\end{equation}
where $B$ denotes an amplitude of the DC transverse fields.
After preparing a ground state of the drive Hamiltonian, we let the state evolve by the annealing Hamiltonian.
If this dynamics is adiabatic, the ground state of the drive Hamiltonian is changed into that of the problem Hamiltonian.

\subsection{Spin lock quantum annealing}

We review the spin-lock quantum annealing~\cite{chen2011experimental,nakahara2013lectures,matsuzaki2020quantum}.
Unlike the conventional QA with the DC transverse fields, we use AC driving fields for the drive Hamiltonian when we adopt the spin-lock QA.
Since the previous 
studies focus only on
Ising interaction for the problem Hamiltonian, we consider such a case here.
In the case of the spin lock quantum annealing, the annealing Hamiltonian is given by
\begin{equation}
    H(\lambda)=B\sum_{j=1}^{L}\cos(\omega t)\hat{\sigma}_{j}^{(x)}+\sum_{j=1}^{L}\frac{\omega}{2}\hat{\sigma}_{j}^{(z)}+(1-\lambda)H_{P}\label{eq:spinlockqa_hamiltonian}
\end{equation}
where $B$ denotes an amplitude of AC field,
$\hat{\sigma}_{j}^{(k)}(k=x,y,z)$ denotes the Pauli operator acting on the $j$-th site of the spin chain, $L$ is the site number,
and $H_{P}$ is the problem Hamiltonian. 
By moving to a rotating frame defined by $U$, the Hamiltonian is rewritten as
\begin{equation}
    H'=UHU^{\dag}-iU^{\dag}\frac{dU}{dt}
\end{equation}
Since we drive the qubits with the AC driving fields, we choose the operator $U$ as follows
\begin{equation}
    U=\exp\biggl(-i\frac{\omega}{2}t\sum_{j=1}^{L}\hat{\sigma}_{j}^{(z)}\biggr).
\end{equation}
Here, we have
\begin{equation}
U^{\dag}H_{P}U=H_{P}
\end{equation}
and we can calculate as follows.

\begin{eqnarray}
    U\cos({\omega t})\hat{\sigma}_{j}^{(x)}U^{\dag}&=&\frac{1}{2}\Bigl((e^{2i\omega t}+1)\hat{\sigma}_{j}^{(+)}+(e^{-2i\omega t}+1)\hat{\sigma}_{j}^{(-)}\Bigr)\nonumber\\
    &\approx&\frac{1}{2}\hat{\sigma}_{j}^{(x)}.
\end{eqnarray}
where we use the rotating wave approximation. Throughout of this paper, we use the notation $\approx$ to represent the rotating wave approximation.
Using the relation
\begin{equation}
    -iU^{\dag}\frac{dU}{dt}=-\frac{\omega}{2}\sum_{j=1}^{L}\hat{\sigma}_{j}^{(z)},
\end{equation}
we obtain the annealing Hamiltonian in the rotating frame as follows.
\begin{equation}
    H'\approx\sum_{j=1}^{L}\frac{\lambda}{2}\hat{\sigma}_{j}^{(x)}+(1-\lambda)H_{P}
\end{equation}
Therefore, we obtain the annealing Hamiltonian where the drive Hamiltonian is the transverse field in the rotating frame.
It is worth mentioning that a violation of the the rotating wave approximation due to strong driving could cause an error in the preparation of the ground state~\cite{matsuzaki2020quantum}.

\section{Realization of the XXZ model
using flux qubit}\label{sec:method}
Here, we describe our scheme to prepare a ground state of the Hamiltonian with Ising interaction and energy-exchange interaction by using the spin-lock QA.
Here, we consider a case that the flux qubits are inductively coupled.
The Hamiltonian of the flux qubit~\cite{friedman2000quantum,van2000quantum,clarke2008superconducting} or capacitively shunted FQs~\cite{yan2016flux}
is given by
\begin{equation}
    H=\frac{\epsilon}{2}\sum_{j=1}^{L}\hat{\sigma}_{z}^{(j)}+\frac{\Delta_{\rm{G}}}{2}\sum_{j=1}^{L}\hat{\sigma}_{x}^{(j)}
    +\lambda\cos{\omega t}\sum_{j=1}^{L}\hat{\sigma}_{y}^{(j)}
    +g\sum_{j=1}^{L-1}\hat{\sigma}_{z}^{(j)}\hat{\sigma}_{z}^{(j+1)}.
\end{equation}
where $\epsilon$ denotes an energy bias, $\Delta_{\rm{G}}$ denotes an energy gap, $g$ denotes a strength of the inductive interactions, and $\lambda$ denotes a Rabi frequency. 
First, we use the single qubit operator $U_{y}\equiv e^{-i\theta\sum_{j=1}^{L}\hat{\sigma}_{y}^{(j)}}$ to rotate this Hamiltonian by $\theta$ in the y-direction.
\begin{eqnarray}
    H'&=&U_{y}^{-1}HU_{y}\nonumber\\
    &=&\biggl(\frac{\Delta_{\rm{G}}}{2}\cos{\theta}-\frac{\epsilon}{2}\sin{\theta}\biggr)\sum_{j=1}^{L}\hat{\sigma}_{x}^{(j)}
    +\biggl(\frac{\epsilon}{2}\cos{\theta}+\frac{\Delta_{\rm{G}}}{2}\sin{\theta}\biggr)\sum_{j=1}^{L}\hat{\sigma}_{z}^{(j)}+\lambda\cos{\omega t}\sum_{j=1}^{L}\hat{\sigma}_{y}^{(j)}\nonumber\\
    &+&g\sum_{j=1}^{L-1}\biggl(\cos^{2}{\theta}\hat{\sigma}_{z}^{(j)}\hat{\sigma}_{z}^{(j+1)}+\sin^{2}{\theta}\hat{\sigma}_{x}^{(j)}\hat{\sigma}_{x}^{(j+1)}-\cos{\theta}\sin{\theta}(\hat{\sigma}_{z}^{(j)}\hat{\sigma}_{x}^{(j+1)}+\hat{\sigma}_{x}^{(j)}\hat{\sigma}_{z}^{(j+1)})\biggr)\nonumber\\
\end{eqnarray}
In order to remove
the term of $\sum_{j=1}^{L}\hat{\sigma}_{x}^{(j)}$, we determine the $\theta$ as follows
\begin{eqnarray}
    \sin{\theta}=\frac{\Delta_{\rm{G}}}{\sqrt{\epsilon^{2}+\Delta_{\rm{G}}^{2}}},\\ 
    \cos{\theta}=\frac{\epsilon}{\sqrt{\epsilon^{2}+\Delta_{\rm{G}}^{2}}}.
\end{eqnarray}
Next, the Hamiltonian in the interaction picture is given by
\begin{eqnarray}
    H''=U^{-1}H'U-iU^{-1}\frac{dU}{dt}
\end{eqnarray}
where the single qubit operator $U$ is defined as 

\begin{eqnarray}
U=e^{-\frac{i\omega t}{2}\sum_{j=1}^{L}\hat{\sigma}_{z}^{(j)}}.
\end{eqnarray}
Using the relation 
\begin{eqnarray}
    -iU^{-1}\frac{dU}{dt}=-\frac{\omega}{2}\sum_{j=1}^{L}\hat{\sigma}_{z}^{(j)}\nonumber,
\end{eqnarray}
we obtain
\begin{eqnarray}
    H''&=&\lambda\cos{\omega t}U^{-1}\sum_{j=1}^{L}\hat{\sigma}_{y}^{(j)}U
    +\biggl(\frac{\epsilon}{2}\cos{\theta}+\frac{\Delta_{\rm{G}}}{2}\sin{\theta}-\frac{\omega}{2}\biggr)U^{-1}\sum_{j=1}^{L}\sigma_{z}^{(j)}U\nonumber\\
    &+&\sum_{j=1}^{L-1}g\biggl(\cos^{2}{\theta}U^{-1}\hat{\sigma}_{z}^{(j)}\hat{\sigma}_{z}^{(j+1)}U+\sin^{2}{\theta}U^{-1}\hat{\sigma}_{x}^{(j)}\hat{\sigma}_{x}^{(j+1)}U\nonumber\\
    &-&\cos{\theta}\sin{\theta}U^{-1}(\hat{\sigma}_{z}^{(j)}\hat{\sigma}_{x}^{(j+1)}+\hat{\sigma}_{x}^{(j)}\hat{\sigma}_{z}^{(j+1)})U\biggr).\label{eq:H''_annealing_ham}
\end{eqnarray}
We calculate each term of this Hamiltonian.
The first term of this Hamiltonian is  
\begin{eqnarray}
    \lambda\cos{\omega t}U^{-1}\hat{\sigma}_{y}^{(j)}U&=&
\frac{\lambda}{4i}\biggl(e^{2i\omega t}\hat{\sigma}_{+}^{(j)}+\hat{\sigma}_{+}^{(j)}-e^{-2i\omega t}\hat{\sigma}_{-}^{(j)}-\hat{\sigma}_{-}^{(j)}\biggr)\nonumber\\
    &\approx&\frac{\lambda}{2}\hat{\sigma}_{y}^{(j)}\label{eq:lamUsigmayU}
\end{eqnarray}
by rotating
wave approximation.
We obtain the following

\begin{eqnarray}
    U^{-1}\hat{\sigma}_{x}^{(j)}\hat{\sigma}_{x}^{(j+1)}U
    &=&e^{2i\omega t}\hat{\sigma}_{+}^{(j)}\hat{\sigma}_{+}^{(j+1)}+e^{-2i\omega t}\hat{\sigma}_{-}^{(j)}\hat{\sigma}_{-}^{(j+1)}
    +\hat{\sigma}_{+}^{(j)}\hat{\sigma}_{-}^{(j+1)}
    +\hat{\sigma}_{-}^{(j)}\hat{\sigma}_{+}^{(j+1)}\nonumber\\
    &\approx&\hat{\sigma}_{+}^{(j)}\hat{\sigma}_{-}^{(j+1)}
    +\hat{\sigma}_{-}^{(j)}\hat{\sigma}_{+}^{(j+1)}\label{eq:UsigmaxsigmaxU}.
\end{eqnarray}
Also, we obtain
\begin{eqnarray}
    U^{-1}(\hat{\sigma}_{z}^{(j)}\hat{\sigma}_{x}^{(j+1)}+\hat{\sigma}_{x}^{(j)}\hat{\sigma}_{z}^{(j+1)})U
    &=&\hat{\sigma}_{z}^{(j)}(e^{i\omega t}\hat{\sigma}_{+}^{(j+1)}+e^{-i\omega t}\hat{\sigma}_{-}^{(j+1)})
    +(e^{i\omega t}\hat{\sigma}_{+}^{(j)}+e^{-i\omega t}\hat{\sigma}_{-}^{(j)})\hat{\sigma}_{z}^{(j+1)}\nonumber\\
    &\approx& 0\label{eq;UsigmazsigmaxU}.
\end{eqnarray}
Finally, the other terms are unaffected by the operator $U$.
Therefore, substituting the Eq. (\ref{eq:lamUsigmayU}), (\ref{eq:UsigmaxsigmaxU}) and, (\ref{eq;UsigmazsigmaxU}) into the Eq. (\ref{eq:H''_annealing_ham}), we obtain the 
\begin{eqnarray}
    H''\approx\frac{\lambda}{2}&\sum_{j=1}^{L}\hat{\sigma}_{y}^{(j)}+\Delta\omega\sum_{j=1}^{L}\hat{\sigma}_{z}^{(j)}\nonumber\\
    &+g\sum_{j=1}^{L-1}\biggl(\frac{\epsilon^{2}}{\epsilon^{2}+\Delta_{\rm{G}}^{2}}\hat{\sigma}_{z}^{(j)}\hat{\sigma}_{z}^{(j+1)}+\frac{\Delta_{\rm{G}}^{2}}{\epsilon^{2}+\Delta_{\rm{G}}^{2}}(\hat{\sigma}_{+}^{(j)}\hat{\sigma}_{-}^{(j+1)}+\hat{\sigma}_{-}^{(j)}\hat{\sigma}_{+}^{(j+1)})\biggr)\ \ \ \ \ \ \ \ \ \ \ \ \ 
    \label{eq:ann_ham+spinlock}
\end{eqnarray}
where the detuning is defined as
$\delta\omega=\biggl(\frac{\epsilon^{2}}{2\sqrt{\epsilon^{2}+\Delta_{\rm{G}}^{2}}}+\frac{\Delta_{\rm{G}}^{2}}{2\sqrt{\epsilon^{2}+\Delta_{\rm{G}}^{2}}}-\frac{\omega}{2}\biggr)$.
By rotating the Hamiltonian by $\pi/2$ along the z axis, we obtain the Hamiltonian with the transverse field parallel to the x axis.
This means that, by changing $\lambda$ and $g$, we can perform the QA where the driver Hamiltonian is the transverse field and the problem Hamiltonian is the XXZ model.
Throughout of our paper, we take $\lambda (t)=\lambda(0) \frac{t}{T}$ and $g (t)=g(0) (1- \frac{t}{T})$.
Moreover, since we have not only the transverse field parallel to the x axis but also the longitudinal field, we could construct a driver Hamiltonian with twisted fields, which is shown to be useful to obtain a ground state with the QA \cite{imoto2021quantum,kadowaki2021greedy}.
However, in our paper, we focus on the resonant case with $\delta \omega =0$.

\section{Application and numerical simulation}\label{sec:numerical_simulation}

In this section, in order to evaluate the performance of our scheme, we perform  numerical simulations. 
For the problem Hamiltonians, we consider an example of a two dimensional anisotropic Heisenberg model in condensed matter physics.

\begin{figure}[ht]
    \centering
    \includegraphics[width=65mm]{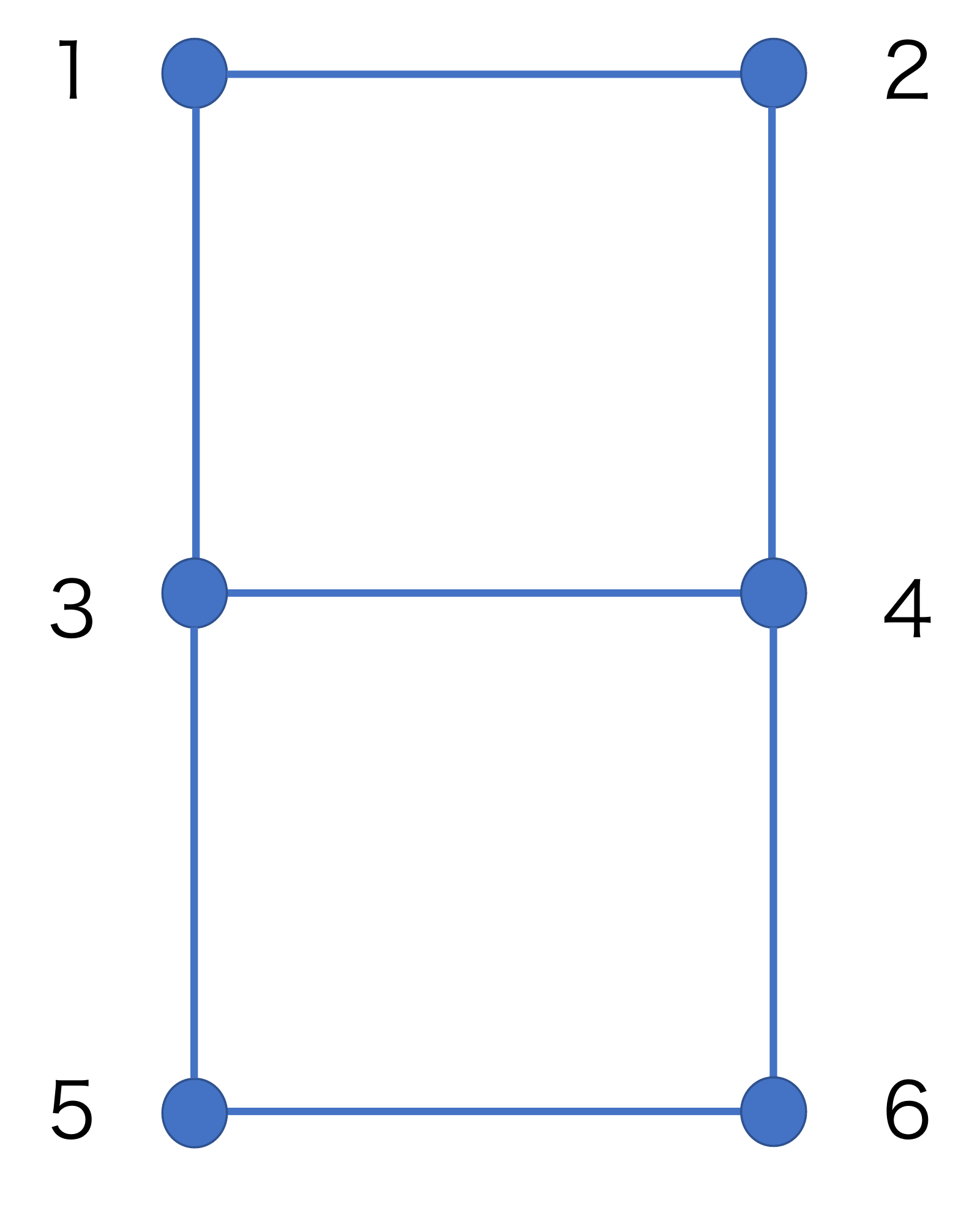}
    \caption{Schematic of our model.
    }
    \label{fig:square_lattice}
\end{figure}

\begin{figure}[ht]
    \centering
    \includegraphics[width=150mm]{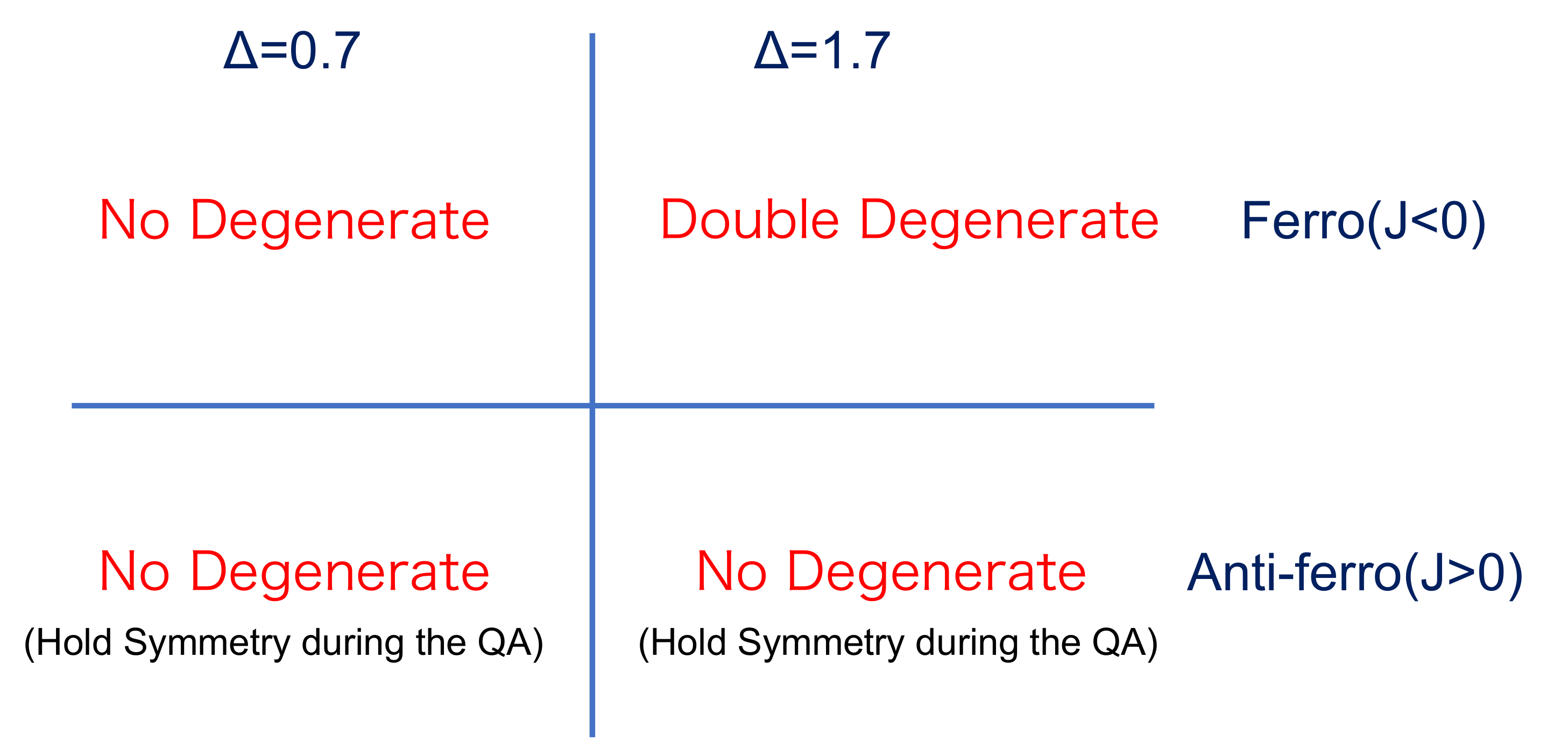}
    \caption{
    A conceptual diagram of the anisotropic Heisenberg model.}
    \label{fig:heisenberg_table}
\end{figure}

The problem Hamiltonian is given by 
\begin{equation}
    H_{\rm{P}}=J\sum_{\braket{j,k}}\biggl(\sigma_{j}^{(x)}\sigma_{k}^{(x)}+\sigma_{j}^{(y)}\sigma_{k}^{(y)}+\Delta\sigma_{j}^{(z)}\sigma_{k}^{(z)}\biggr).
\end{equation}
where $\Delta$ denotes the anisotropic parameter, and $J$ denotes the coupling constant.
This is called the ferro magnetic model when $J<0$, and anti-ferro magnetic model when $J>0$.
To realize this problem Hamiltonian in the spin lock we set $J=g\frac{\epsilon^{2}}{\epsilon^{2}+\Delta_{\rm{G}}^{2}}$ and  $\Delta=\frac{\Delta_{\rm{G}}^{2}}{\epsilon^{2}}$.
We consider a  square lattice of size $2\times3$, as shown in Figure.\ref{fig:square_lattice}. 
About the coupling constants $J$ and anisotropic parameters $\Delta$, we consider the following
four cases.
\begin{equation}
    (J, \Delta)= (1, 1.7), (1, 0.7,), (-1, 1.7,), (-1, 0.7).
\end{equation}
where the unit of the value of $J$ is GHz, as shown in Figure.\ref{fig:heisenberg_table}. 

The ground state was found to be doubly degenerate in the case of ferro-magnetic coupling ($J=-1$) and $\Delta=1.7$. 
On the other hand, the ground state is not degenerate in the other cases.
Throughout of our manuscript, when we plot a fidelity $F$,
we calculate $F=\sum _{n=1}^m \langle g_n | \rho (t) |g_n\rangle $ where $m$ denotes the degeneracy, $\{|g_n\rangle \}_{n=1}^m$ denotes the ground state, and $\rho (t)$ denotes a quantum state after the QA.

\begin{figure}[h]
 \begin{minipage}[b]{0.45\linewidth}
  \centering
  \includegraphics[keepaspectratio, scale=0.45]
  {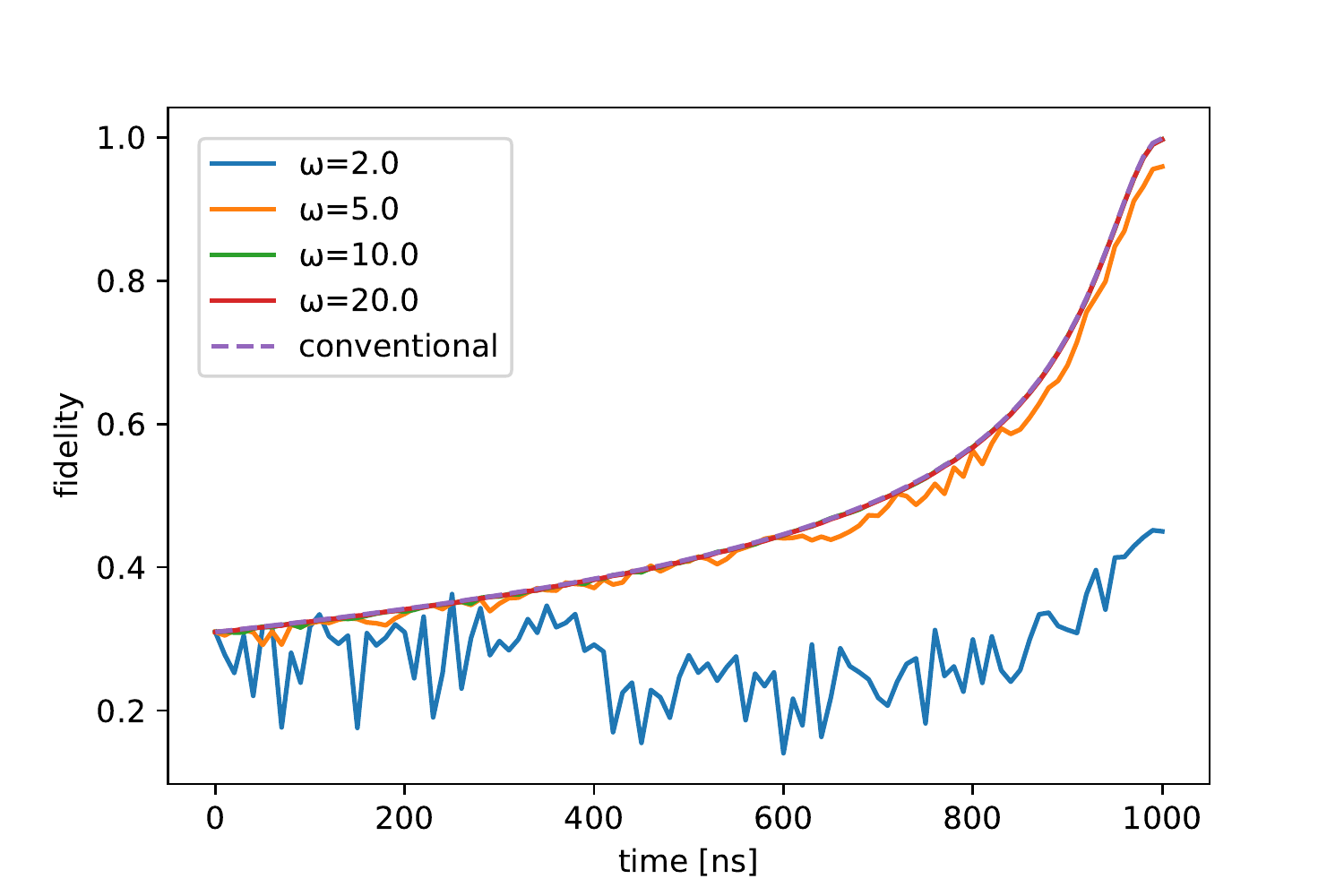}
  \subcaption{$J=-1$ and $\Delta=0.7$}
  \label{j=-/0.7}
 \end{minipage}
 \begin{minipage}[b]{0.45\linewidth}
  \centering
  \includegraphics[keepaspectratio, scale=0.45]
  {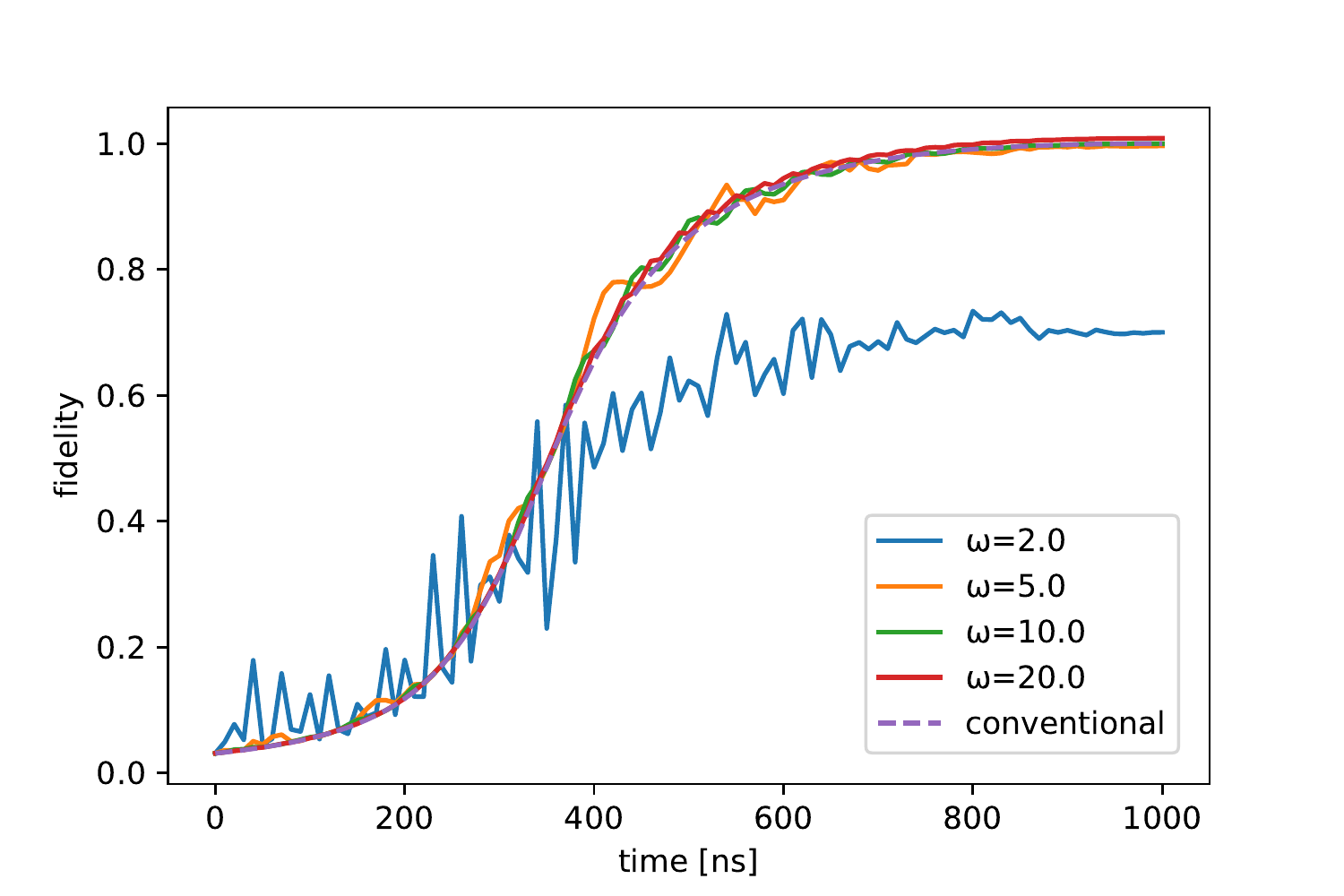}
  \subcaption{$J=-1$ and $\Delta=1.7$}
  \label{j=-/1.7}
 \end{minipage}\\
 \begin{minipage}[b]{0.45\linewidth}
  \centering
  \includegraphics[keepaspectratio, scale=0.45]
  {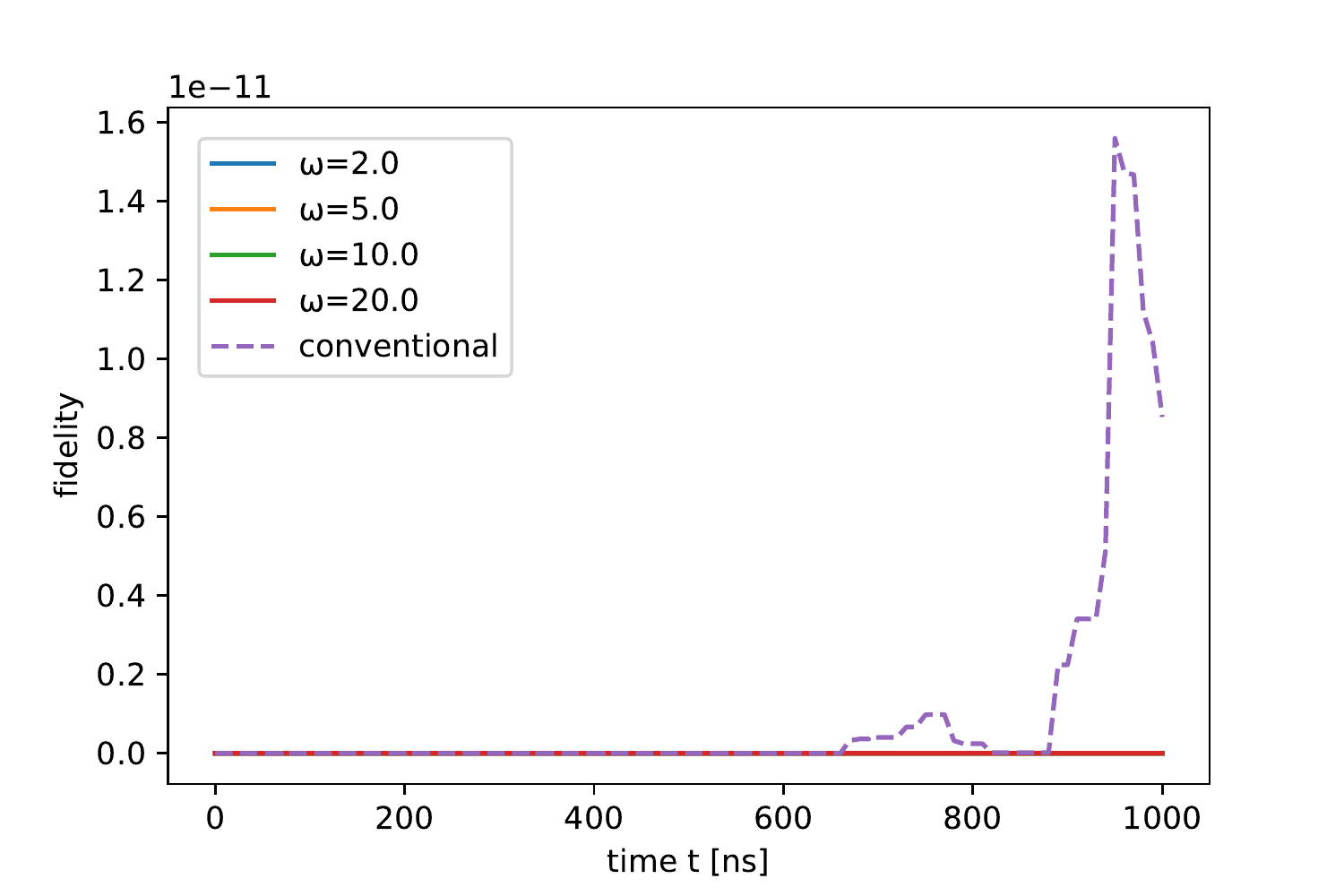}
  \subcaption{$J=1$ and $\Delta=0.7$}
  \label{j=+/0.7}
 \end{minipage}
 \begin{minipage}[b]{0.45\linewidth}
  \centering
  \includegraphics[keepaspectratio, scale=0.45]
  {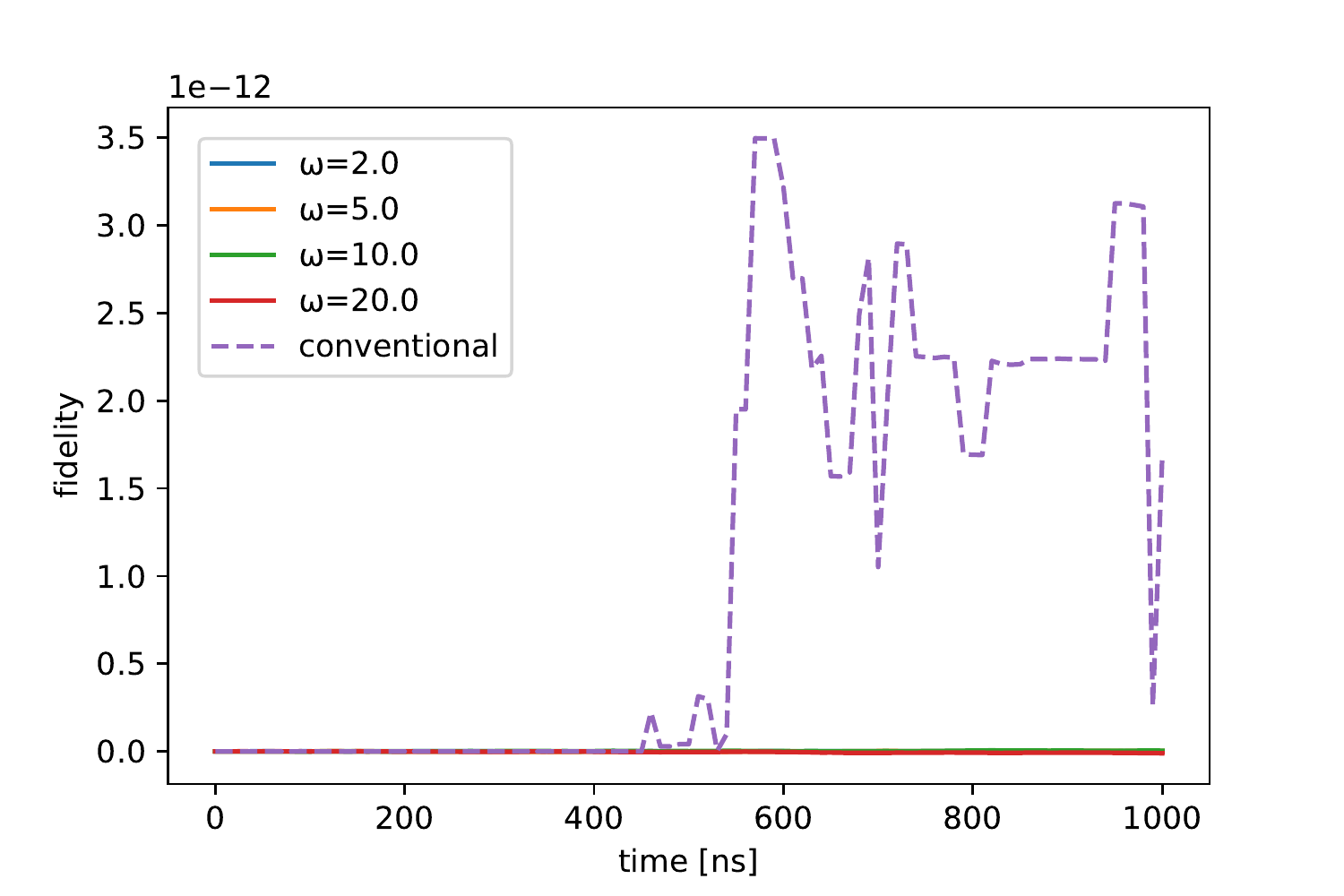}
  \subcaption{$J=1$ and $\Delta=1.7$}
  \label{j=+/1.7}
 \end{minipage}
 \caption{ 
 Fidelity between the ground state of the two-dimensional anisotropic Heisenberg model and the state during QA
 against time.
    The parameters for the simulation are chosen as
    $T=1000$ ns, $\lambda=1$ and, $\omega=2.0, 5.0, 10.0, 20.0$ GHz.
    We adopt the uniform microwave driving (DC transverse fields) as the drive Hamiltonian for our (conventional) scheme.
    }\label{fig:2dHeisenberg_fidelity}
\end{figure}

We plot the fidelity against time for each frequency $\omega$ with numerical simulations in the Figure \ref{fig:2dHeisenberg_fidelity}. 
For these simulations, the annealing time $T$ is set to be $1000$ [ns], the driving strength $\lambda$ is set to be 1 GHz, and the frequencies $\omega$ are set to be $2.0$, $5.0$, $10.0$, $20.0$ GHz.
For the ferro magnetic case, as we increase the frequency $\omega$, the fidelity increases. This is consistent with the fact that the rotating wave approximation becomes more accurate as the qubit frequency increases.
Thus, after the preparation of the ground state with a high fidelity, we can measure an arbitrary correlation function for the ground state of the Hamiltonian, which is useful in the condensed matter physics.

In the case of anti-ferro magnetic ($J=1$ GHz), for $\Delta=0.7$ and $\Delta=1.7$, the fidelities are zero up to numerical errors.
Our considerations suggest that there is a symmetry during QA for these parameters.
In general, when there is a symmetry, the total Hamiltonian commutes with some observables. 
In this case, the Hamiltonian can be block diagonalized into sectors
by applying a suitable unitary operator. 
As long as the initial state belongs to a different sector from that to which the target ground state belongs, we cannot prepare a ground state even after the QA by keeping the adiabatic condition because a level crossing occurs.
Actually, in our case, we confirm that an operator $U_{swap}^{(1,2)}U_{swap}^{(3,4)}U_{swap}^{(5,6)}$ commutes with both the transverse field and the ferromagnetic XXZ model Hamiltonian where $U_{swap}^{(j,k)}$ denotes a swap gate between the $j$-th qubit and the $k$-th qubit.
We calculate the sector of $U_{swap}^{(1,2)}U_{swap}^{(3,4)}U_{swap}^{(5,6)}$ for the ground states of the transvers field and the ferromagnetic XXZ model, and we obtain $\bra{E_{0}^{(TF)}}U_{swap}^{(1,2)}U_{swap}^{(3,4)}U_{swap}^{(5,6)}\ket{E_{0}^{(TF)}}=1$ and $\bra{E_{0}^{(XXZ)}}U_{swap}^{(1,2)}U_{swap}^{(3,4)}U_{swap}^{(5,6)}\ket{E_{0}^{(XXZ)}}=-1$ where $\ket{E_{0}^{(TF)}}$ denotes the ground state of the transvers field and $\ket{E_{0}^{(XXZ)}}$ denotes the ground state of the XXZ model.
Therefore, we can
confirm that  $\ket{E_{0}^{(TF)}}$ and $\ket{E_{0}^{(XXZ)}}$ belong to different sectors, and this is the reason why the QA fails in our numerical simulations.

\begin{table}[htb]
\centering
    \begin{tabular}{|c||c|}
    \hline
      $J_{1}$ & $-0.9052919617126958$ \\
      $J_{2}$ & $0.2500243810835503$ \\
      $J_{3}$ & $-1.931378367720707$ \\
      $J_{4}$ & $-0.007622719480759765$ \\ 
      $J_{5}$ & $-1.259154537693434$ \\ 
      $J_{6}$ & $-1.261510983981174$ \\ \hline
    \end{tabular}
  \caption{Coefficients of random transverse field $\{J_{j}\}_{j=1}^{6}$ used in Fig. \ref{fig:2dHeisenberg_fidelity_random_tf}. The unit of these values is GHz, as described in the main text.}
  \label{tb:rand_tf_coefficient}
\end{table}

\begin{figure}[h]
 \begin{minipage}[b]{0.45\linewidth}
  \centering
  \includegraphics[keepaspectratio, scale=0.45]
  {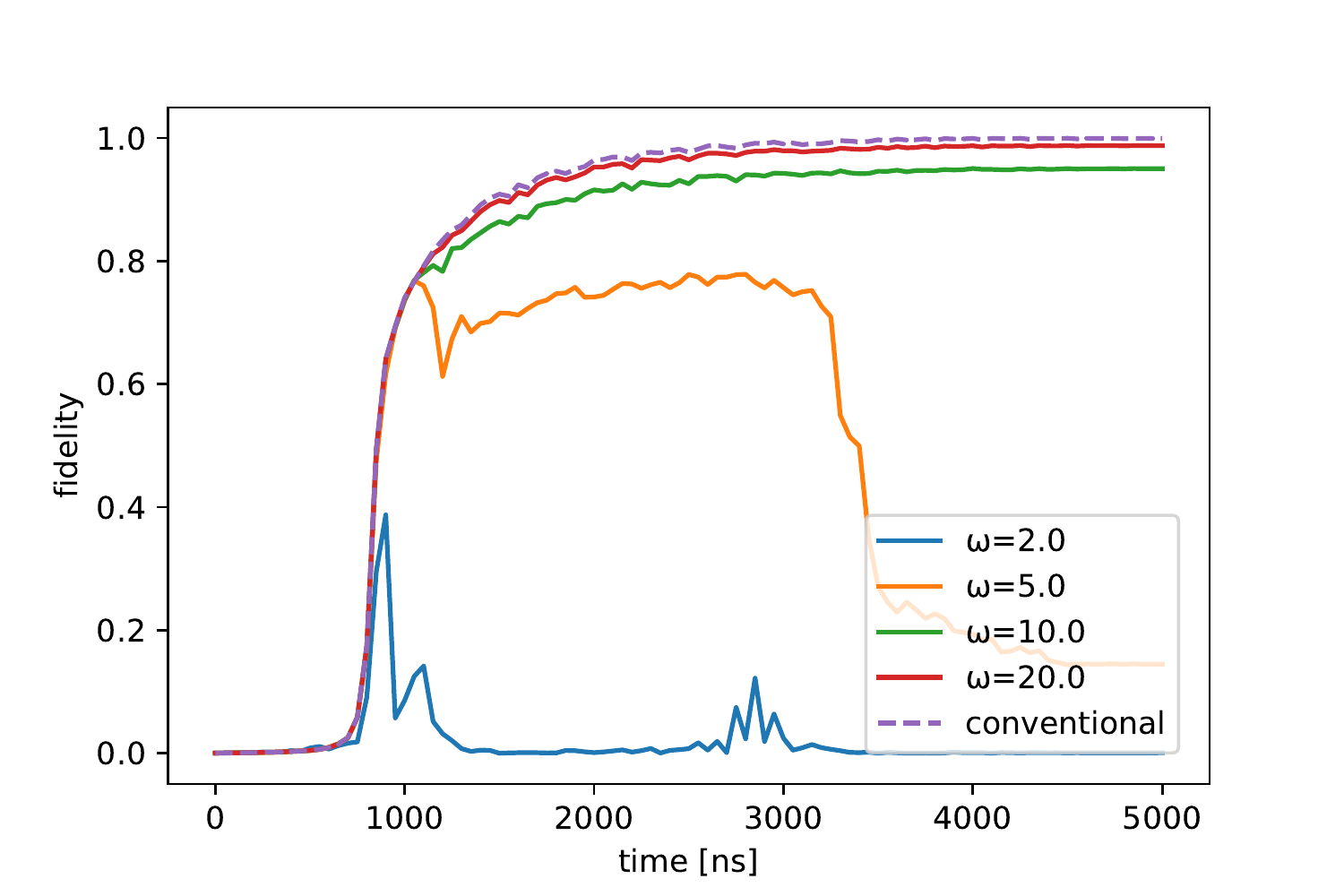}
  \subcaption{$J=1$ and $\Delta=0.7$}
  \label{j=+/0.7}
 \end{minipage}
 \begin{minipage}[b]{0.45\linewidth}
  \centering
  \includegraphics[keepaspectratio, scale=0.45]
  {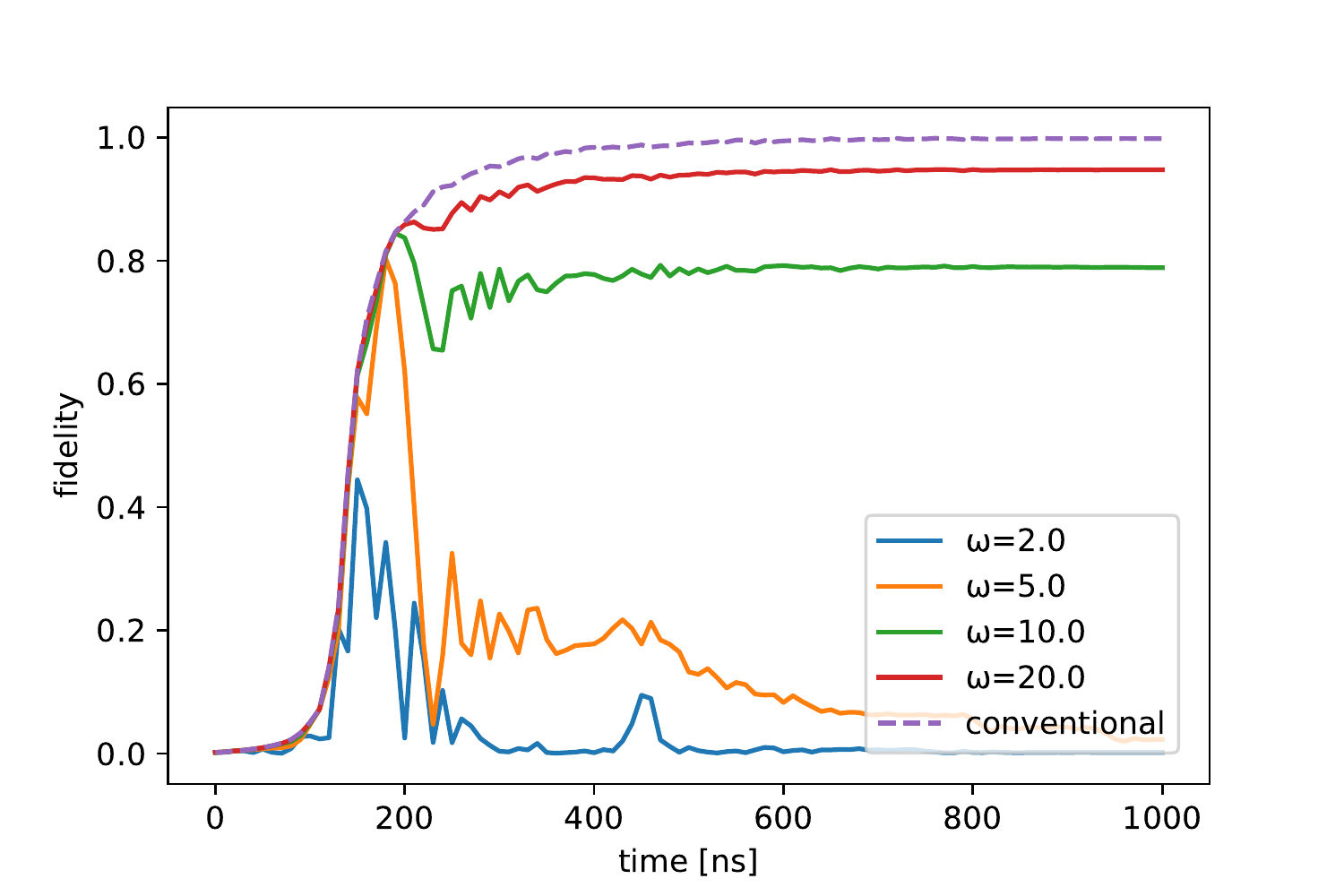}
  \subcaption{$J=1$ and $\Delta=1.7$}
 \end{minipage}
 \caption{We plot the fidelity between the ground state of the two-dimensional anisotropic Heisenberg model and the state during QA against time.
     The parameters for the simulation are chosen as
    , $T=1000$ ns, and $\omega=2.0, 5.0, 10.0, 20.0$ GHz. In order to break the symmetry, we adopt the random 
    microwave driving (DC transverse field) as the drive Hamiltonian for our (conventional) scheme.
    }\label{fig:2dHeisenberg_fidelity_random_tf}
\end{figure}

In such a case, we apply not homogeneous but randomized microwave driving (DC transverse fields) to break the symmetry for our (conventional) scheme. 
Moreover, we plot the fidelities against time when we apply the randomized driving Hamiltonian in the Figure. \ref{fig:2dHeisenberg_fidelity_random_tf}.
This demonstrates that we can prepare a ground state with a high fidelity
by using the randomized fields during QA wihtout level crossing.
Also, our results show that a proper choice of the driver Hamiltonian to break a symmetry of the Hamiltonian is important in QA, which has been overlooked in previous works.

\section{Conclusion}\label{sec:conclusion}
In conclusion, we propose the way to
prepare a ground state of the XXY model with the QA by using inductively coupled superconducting
flux qubits.
The key idea is to use a recently proposed spin-lock quantum annealing. We drive the flux qubits by the microwave driving, and we can construct the
Hamiltonian of QA for the XXY model in the rotating frame.
By choosing the problem Hamiltonian as the anisotropic Heisenberg model,
we can 
prepare a ground state of 
the anisotropic Heisenberg model with our scheme unless there is a level crossing.
Importantly, when there is a symmetry during QA,  the Hamiltonian can be block diagonalized into sectors.
 As long as the initial state belongs to a different sector from that to which the target ground state belongs, we cannot prepare the ground state with QA because of the level crossing. In this case, we show that, by adding randomized fields to break the symmetry, we can avoid the problem of the level crossing.
Our numerical simulation shows that our spin-lock QA actually provides a practical way to prepare a ground state. Our scheme is useful
to know the correlation function of the relevant Hamiltonians in the condensed matter physics.

{\it{Note added.}}—While preparing our manuscript,
we became aware of a related work that also proposes an adiabatic
scheme to prepare a ground state of a Hamiltonian (that contains non-diagonal terms) by applying additional fields to break the symmetry \cite{francis2021determining}.

\ack

We thank S. Kawabata and A. Yoshinaga for a useful advice.
This work was supported by MEXT’s Leading Initiative for Excellent Young Researchers
and JST PRESTO (Grant No. JPMJPR1919), Japan. This paper is partly based on
the results obtained from a project, JPNP16007, commissioned by the New Energy and
Industrial Technology Development Organization (NEDO), Japan.

\appendix

\section*{References}

\bibliography{main}
\end{document}